\documentclass[%
 reprint,
 amsmath,amssymb,
 aps,
 superscriptaddress,
]{revtex4-2}

\usepackage{graphicx}
\usepackage{bm}
\usepackage{braket}
\usepackage{upgreek}
\usepackage{float}
\usepackage[dvipsnames,svgnames,x11names]{xcolor}
\usepackage{ulem}
\usepackage[colorinlistoftodos,prependcaption,textsize=tiny]{todonotes} 
\usepackage{tabularray}

\usepackage{comment}
\usepackage{subfigure}
\usepackage{nicefrac}

\usepackage[pdftex]{hyperref}
\hypersetup{
    breaklinks=true,
    colorlinks=true,
    linkcolor=MidnightBlue,
    urlcolor=NavyBlue,
    citecolor=Blue
    }

\newcommand*\w{{\textrm{w}}}

\begin{document}

\title{Simultaneous measurement of thermal conductivity and specific heat in quasi-2D membranes by $3\omega$ thermal transport}


\author{Yiwei Le}
\affiliation{Department of Physics, Washington University in St. Louis, 1 Brookings Dr., St. Louis MO 63130, USA}
\author{Erdong Song}
\affiliation{Department of Physics, Washington University in St. Louis, 1 Brookings Dr., St. Louis MO 63130, USA}
\author{Jason Li}
\affiliation{Department of Physics, Washington University in St. Louis, 1 Brookings Dr., St. Louis MO 63130, USA}
\author{Erik A.\ Henriksen}
\affiliation{Department of Physics, Washington University in St. Louis, 1 Brookings Dr., St. Louis MO 63130, USA}
\affiliation{Institute of Materials Science \& Engineering, Washington University in St. Louis, 1 Brookings Dr., St. Louis MO 63130, USA}
\email{henriksen@wustl.edu}

\date{\today}

\begin{abstract}
Toward measuring the thermal properties of exfoliated atomically thin materials, we demonstrate  simultaneous measurements of the thermal conductivity and specific heat in suspended membranes. We use the $3\omega$ technique applied to quasi-two-dimensional silicon nitride membranes having a metal line heater patterned ·on the surface to both deliver heat and directly measure the thermal impedance of the membrane at the heating frequency, $Z(2\omega)$. We derive an expression for the complex thermal impedance as a function of frequency, approximating the actual rectangular membranes with a one dimensional model. The derivation accounts for potential parasitic heat loss mechanisms including conduction along the heater line, and by the gas load in an imperfect vacuum. Qualitatively, the thermal impedance response resembles a low-pass filter, owing to the combination of the total thermal resistance and total specific heat. Fitting $Z(2\omega)$ to measurements across a few decades in frequency, we extract values of the thermal conductivity and specific heat of silicon nitride in agreement with literature values. We also study the dependence on the heating current, and compare to measurements of the thermal conductivity at zero frequency.
\end{abstract}



\maketitle

\section{Introduction}

The thermal conductivity and specific heat of materials convey a wealth of information about material properties. Yet such measurement approaches have played a rather small role in the exploration of atomically thin materials over the past two decades. Graphene, as the prototypical 2D material \cite{novoselov_two-dimensional_2005}, has been the focus of a variety of approaches to measure thermal conductivity \cite{Balandin2008,Faugeras2010,Malekpour2017,Li2017,Seol2010,Wang2010,Pettes2011,Xu2014,jang_thickness-dependent_2010,fong_measurement_2013,Crossno2016,Waissman2021,Talanov_2025}, with only a handful of measurements on other materials \cite{Sahoo2013,Jo2013,Jo2014,Lee2015,Yan2013,Judek2015,Zhou2014,Yan2014,Peimyoo2014,Zhang2015,Gu2016,Song2018,Kasirga2020,Kim2021,Kalantari2022}. The specific heat of atomically thin materials is surprisingly underexplored \cite{fong_ultrasensitive_2012,fong_measurement_2013,Li2017,Aamir2021,Liu2024}.

There are practical difficulties in measuring thermal properties of atomically thin materials:\ since ${\sim}10$-$\mu$m-sized, nm-thick exfoliated flakes have the aspect ratio of a typical sheet of kitchen plastic wrap, if heat is injected into such a sample while supported on a substrate, the heat will largely just flow down into the support. Even for the most thermally insulating materials---aerogels \cite{Thapliyal2014}, or turbostratically disordered transition metal dichalcogenides \cite{Kim2021}---the thermal conductance from the membrane to the substrate will greatly outweigh the heat flow through the vanishingly small cross section of the sheet. An obvious alternative---suspending atomically thin samples to eliminate parasitic loss to the substrate---poses difficulties for device fabrication and measurement, but can in principle provide perfect thermal isolation for the bulk of the membrane.

To address these opportunities and challenges, we are adapting the versatile $3\omega$ thermal transport technique \cite{birge_specific-heat_1986,cahill_thermal_1987,cahill_thermal_1990,dames_1omega_2005,dames_measuring_2013} for use with suspended exfoliated samples. Here we develop the experimental and analytical approaches. Since the measurement is intended to apply to any suspended membrane, we initially substitute suspended silicon nitride membranes for exfoliated atomically thin materials, as these membranes are readily available in sizes matching the relevant aspect ratio for exfoliated atomically thin materials (${\sim}10\times10~\mu$m flakes and ${\sim}$nm thick, corresponding to silicon nitride membranes 1 mm on a side and 100 nm thick), and allow the method to be developed apart from fabrication challenges of exfoliated materials. 

Toward a robust method for directly accessing the thermal properties of atomically thin materials, we choose the $3\omega$ technique over alternative approaches---optical\cite{Balandin2008,Faugeras2010,Yan2013,Yan2014,Zhou2014,Peimyoo2014,Judek2015,Zhang2015,Malekpour2017,Li2017}, direct measurement of thermal gradients \cite{jang_thickness-dependent_2010}, microbridges \cite{Seol2010,Wang2010,Pettes2011,Jo2013,Xu2014,Jo2014,Lee2015}, high frequency noise (accessing only the electronic component) \cite{fong_ultrasensitive_2012,fong_measurement_2013,Crossno2016,Aamir2021,Waissman2021,Talanov_2025}, or nanomechanical resonators \cite{Liu2024}---for two reasons:\ First, and most importantly, in contrast to these methods, the $3\omega$ technique applied to suspended membranes can access \textit{both} the thermal conductivity and the specific heat simultaneously.


Second, the $3\omega$ technique has straightforward and convenient requirements on device patterning and measurement. Only a single metallic wire patterned on the material of interest is required in order to directly measure the thermal impedance at the heating frequency, $Z(2\omega)$. This quantity directly encodes the thermal properties of the material that impact heat transport from the wire. The measurement entails routine lock-in detection of the 3$^{rd}$ harmonic of the ac electrical current applied to the wire; a calibration of the wire resistance vs temperature; and control over the sample environment. No differential thermometry is required. Finally this approach is applicable to most any material whether electrically conducting, insulating, or magnetic; and is compatible with low-temperature experiments, high magnetic fields, and arbitrary sample stage rotations, all of which make competing measurement methods (optical approaches in particular) a challenge.


\begin{figure}[t!]
    \includegraphics[width=\columnwidth]{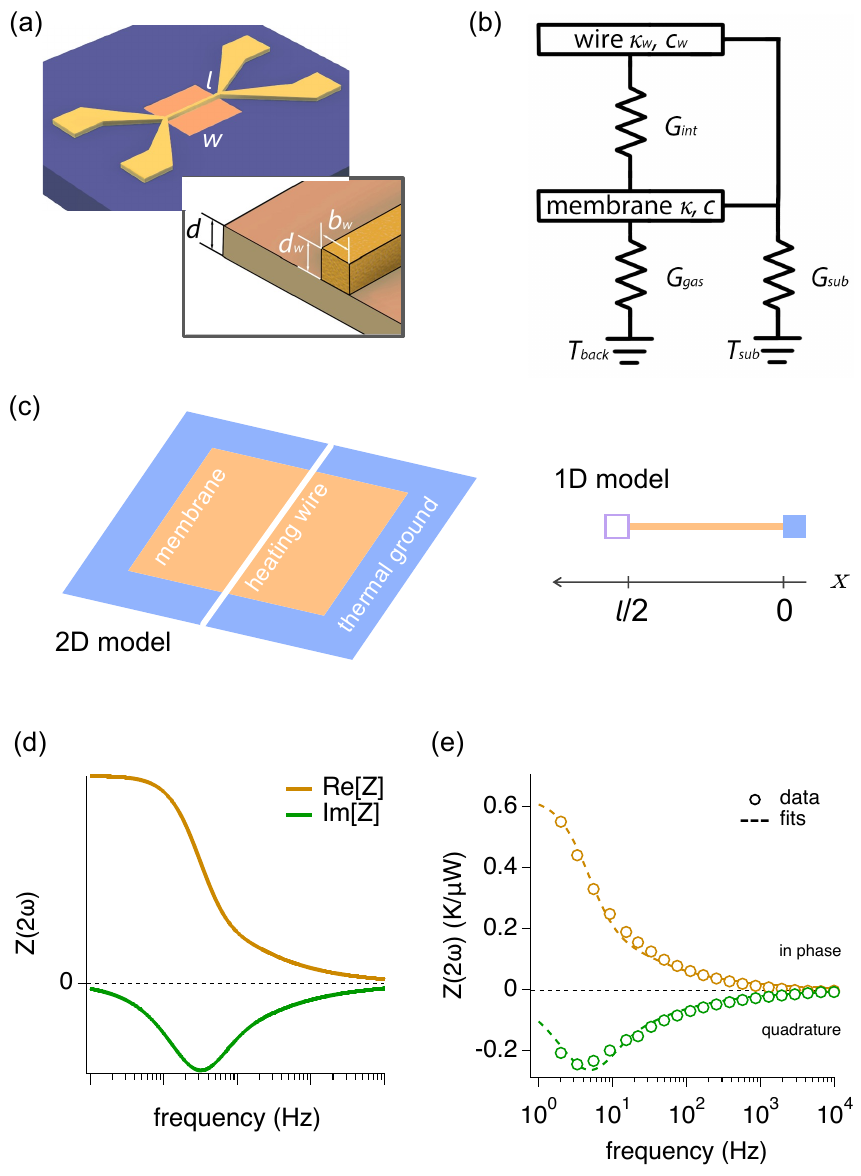}
    \caption{(a) Schematic of device: suspended rectangular membrane (orange, dimensions $l{\times}$\w) on thermally conducting substrate (blue) with electrical connections and heating wire on membrane, (gold). Inset shows membrane thickness $d$ and heating wire dimensions $d_\w$ and $b_\w$. (b) Thermal impedance model including thermal conductivity and specific heat of the membrane and wire, and relevant thermal conductances $G_{int}$ between the wire and membrane; $G_{gas}$ to the cryostat thermal background; and $G_{sub}$ to the substrate. (c) Left:\ 2D model of membrane, with thermal boundary at the substrate temperature, and heating from a wire. Right:\ 1D model of a rod heated at the left by a thermal mass representing the wire, and cooled to thermal ground at right. (d) Real and imaginary parts of the thermal impedance at the heating frequency, $Z(2\omega)$. (e) In phase and quadrature data for $Z(2\omega)$ of a suspended Si$_3$N$_4$ membrane measured at $T = 200$ K, with fits to the $Z(2\omega)$ formula from which the membrane $\kappa$ and $c$ can be extracted.}\label{3w}
\end{figure}

We begin by reviewing the $3\omega$ measurement, working in a one-dimensional approximation to the quasi-two-dimensional samples. We describe the electronic and thermal transfer functions, including how the thermal impedance, $Z(2 \omega)$, depends on the thermal conductivity, $\kappa$, and specific heat, $c$, of the material supporting the heating wire. We also account for two potential parasitic heat loss mechanisms:\ conduction to the substrate along the heating wire, and via the gas load in a non-ideal vacuum. The formulae obtained are then used to extract $\kappa$ and $c$ for a silicon nitride membrane. Although in this work we do not find any appreciable parasitic thermal losses, we anticipate they may become relevant in future measurements of microscopic samples. 

\section{Theory}

The $3\omega$ measurement uses a single metallic wire patterned on the material whose thermal properties are to be measured. Crucially, the resistance of the wire must have some dependence on temperature, $R\equiv R(T)$. An ac electric current, $I(\omega)$, is passed through the wire, heating it sufficiently out of equilibrium to induce $2\omega$ oscillations of the wire resistance. Mixing of these frequencies yields the desired voltage at the third harmonic, $V_{3\omega}$. This voltage essentially captures how the wire cools via the material it is patterned on. 

The device schematic is shown in Fig.\ \ref{3w}(a), consisting of a quasi-2D membrane of thickness $d$ suspended over a rectangular hole of area $l{\times}\w$. So long as $d<<\lambda$, where $\lambda$ is the thermal wavelength defined below, there is no thermal gradient normal to the plane and thermal transport is in the 2D limit. The membrane has a heating wire of width $b_\w$ and thickness $d_\w$ patterned on it, with multiple electrical leads for a four-terminal measurement. The thermal circuit is shown in Fig.\ \ref{3w}(b), accounting for the wire thermal conductivity $\kappa_\w$ and specific heat $c_\w$, and the thermal conductances $G_{int}$ across the interface to the membrane and $G_{sub}$ to the substrate at thermal ground. The membrane has thermal conductivity $\kappa$ and specific heat $c$---these are the quantities we desire to measure---and the  $G_{gas}$ captures heat lost to (or gained from) the background in a non-ideal vacuum.


The ideal device is shown at left in Fig.\ \ref{3w}(c), a two-dimensional rectangular membrane with edges at thermal ground, and heated by a current passed through the wire. For simplicity, we approximate this system with a one-dimensional model  shown at right in Fig.\ \ref{3w}(c), having a rod with conductivity and specific heat $\kappa$ and $c$ that transports heat from a thermal mass representing the wire, at left, to thermal ground at right.

\subsection*{Electrical and thermal transfer functions}

The substrate around the membrane is held at base temperature $T_0$. An ac current $I(\omega) = I_0 \textrm{sin}(\omega t)$ generates a power $P=I^2 R$ in the wire, increasing its average temperature to $\Delta T = T_{avg} - T_0$. These quantities are related by the thermal impedance, $Z$:
\begin{align}
\Delta T = P * Z \label{conv}
\end{align}
which is a convolution of the responses at relevant heating frequencies. For small powers, the first order change in resistance is given by
\begin{align}
R(T) &= R_0(1 + \beta \Delta T) = R_0 (1+\beta I^2*Z) \nonumber\\
&= R_0 +  \frac{\beta I_0^2 R_0}{2}\left[Z_{0\omega}-Z_{2\omega} \textrm{cos}(2\omega t)\right]
\end{align}
where $\beta \equiv dR/dT$ and we explicitly show the Fourier components of $Z$ in response to the power at dc and $2\omega$. The voltage drop across the wire becomes
\begin{align}
V &= I R \nonumber\\
&=  \left[I_0 R_0 + \frac{\beta I_0^3 R_0^2}{2} \left(Z_0 + \frac{Z_{2\omega}}{2} \right) \right] \textrm{sin}(\omega t) \nonumber \\
&~~~~~~ - \frac{\beta I_0^3 R_0^2}{4} ~Z_{2\omega} \textrm{sin}(3 \omega t) \label{harmonics}
\end{align}
The second term is the desired voltage at the third harmonic, $V_{3\omega}$. Note the Ohmic response at $\omega$ gains a contribution proportional to $I^3$ and both the dc and $2\omega$ thermal impedances; thus care must be taken when calibrating $R(T)$ to use small currents that do not appreciably heat the wire.

Standard low-frequency lockin voltage measurements typically record rms values. Correspondingly, the thermal impedance can be calculated from measured quantities using
\begin{align}
    Z_{2\omega}=-\frac{2V_{3\omega,rms}}{\beta I_{0,rms}^3 R_0^2} \label{zm}
\end{align} 
We derive the dependence of $Z_{2\omega}$ on the material parameters and geometry in the Appendix; the response for heating by a current at frequency $\omega$ is found to be
\begin{align}
Z(2\omega) = \frac{1}{2 \w}\left(d k \kappa\ \textrm{coth}(kl/2) + G_\w/\w + i \omega b_\w c_\w d_\w \right)^{-1} \label{zeq}
\end{align}
Here $G_\w$ is an effective thermal conductance due to heat flowing out the ends of the wire. The wavenumber 
\begin{align}
k = \sqrt{(i 2 \omega + \gamma)/\alpha} \label{keq}
\end{align}
with
\begin{align}
\alpha &= \frac{\kappa d}{c d + c_g d_g}  \label{alphaeq} \\
\gamma &= \frac{2 K_{\perp}}{c d+c_g d_g}  \label{gammaeq}
\end{align}
where $K_{\perp}$ is the surface thermal conductivity. If the thermal conductance of the wire and by any gas load is ignored, these equations simplify somewhat with $G_\w = \gamma = 0$ and $k=\sqrt{i 2 \omega c/\kappa}$. 

The real and imaginary parts of $Z_{2\omega}$ for this 1D model are plotted in Fig.\ \ref{3w}(d). These closely resemble the response of a low pass filter, reflecting how the system can be described by a lumped element model having total thermal resistance $R\sim 1/\kappa $ and specific heat $C\sim c$, in close analogy to an electronic $RC$ filter \cite{dames_1omega_2005}.

Equation \ref{zeq} accurately describes the measured thermal impedance of our SiN device. In Fig.\ \ref{3w}(e) we show an example of the data for $T = 200$ K. The behavior of the in phase and quadrature ($90^{\circ}$ out of phase) parts of the measured impedance do indeed resemble a low pass filter, and can be well-fit by the Re and Im parts of $Z(2\omega)$ in order to extract $\kappa$ and $c$ of the membrane as fit parameters.

\begin{figure}[t!]
    \includegraphics[width=\columnwidth]{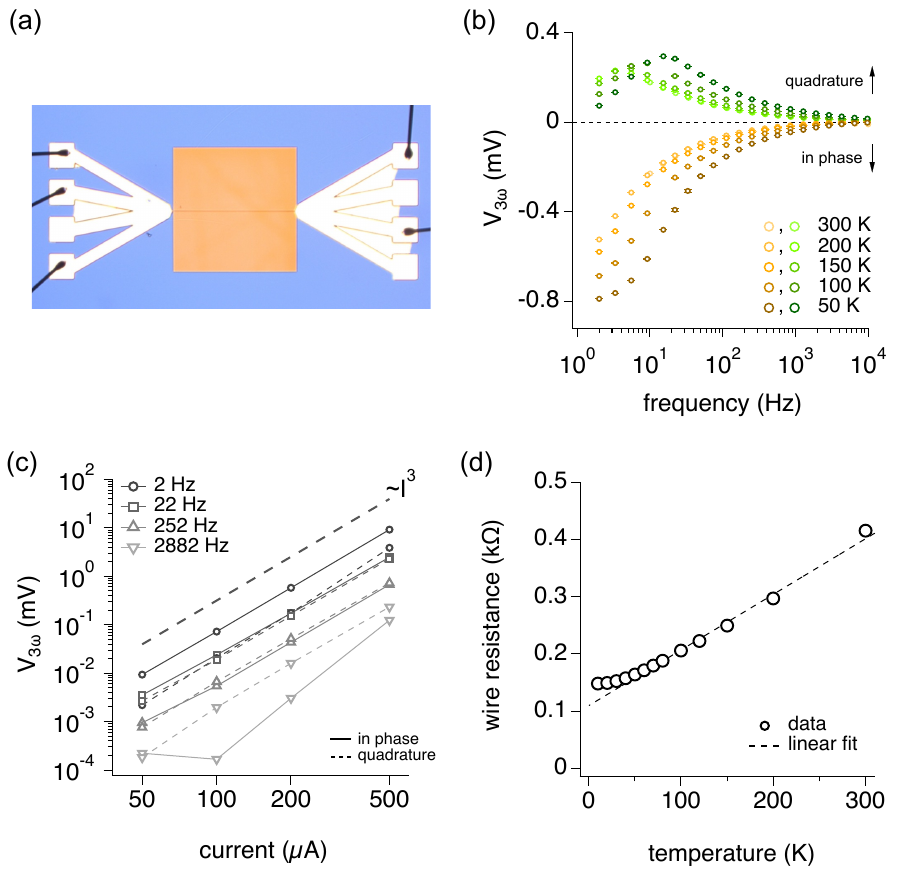}
    \caption{(a) Micrograph of SiN membrane device (100-nm-thick membrane appears orange). (b) In phase and quadrature $V(3\omega)$ voltages vs frequency at various temperatures. (c) Variation of $V(3\omega)$ with current at different frequencies, showing the expected $\propto I_{\omega}^3$ dependence. (d) Four-terminal resistance of the patterned heating wire vs temperature.}\label{dev}
\end{figure}

\section{Experimental Results}

Figure \ref{dev}(a) shows a typical device that closely realizes the schematic of Fig.\ \ref{3w}. It consists of a 100-nm-thick silicon nitride sheet on a $200~\mu$m thick silicon substrate, where the central 1 mm $\times$ 1 mm portion is a suspended membrane on which we pattern a $1~\mu$m wide, 100-nm-thick Au wire contacted by several electrodes. The measurement proceeds by applying an ac signal from a current source, with a lockin used to detect the voltage at the third harmonic, $V(3\omega)$, in a four-terminal configuration. 

In Fig.\ \ref{dev}(b) we show the in phase and quadrature parts of this voltage at various temperatures, measured with a 200 $\mu$A bias. With decreasing temperature, we observe a consistent increase in the magnitude of both parts and an upward shift in frequency of the peak in the quadrature. In separate measurements, we confirm that the voltage at the second harmonic is consistent with zero to within the noise, as expected from Eq.\ \ref{harmonics}. 

The analysis of the preceding section predicts that $V_{3\omega}$ is proportional to the cube of the current. In Fig.\ \ref{dev}(c) we plot the dependence of $V(3\omega)$ on the bias current, which indeed matches expectations except for data at the highest frequency plotted, for which values of $V(3\omega)$ lie close to the noise floor. 

The resistance $R(T)$ of the wire is needed to calculate $Z(2\omega)$. Figure \ref{dev}(d) shows $R(T)$ measured using a small bias current ($I(\omega) = 1~\mu$A) to avoid self-heating. As is typical for an evaporated thin film of a nominally pure metal, the resistance saturates at low temperatures and otherwise increases linearly with  temperature, with slope $\beta = 1.00(0.03)~\Omega$/K.

\begin{figure}[t!]
    \includegraphics[width=\columnwidth]{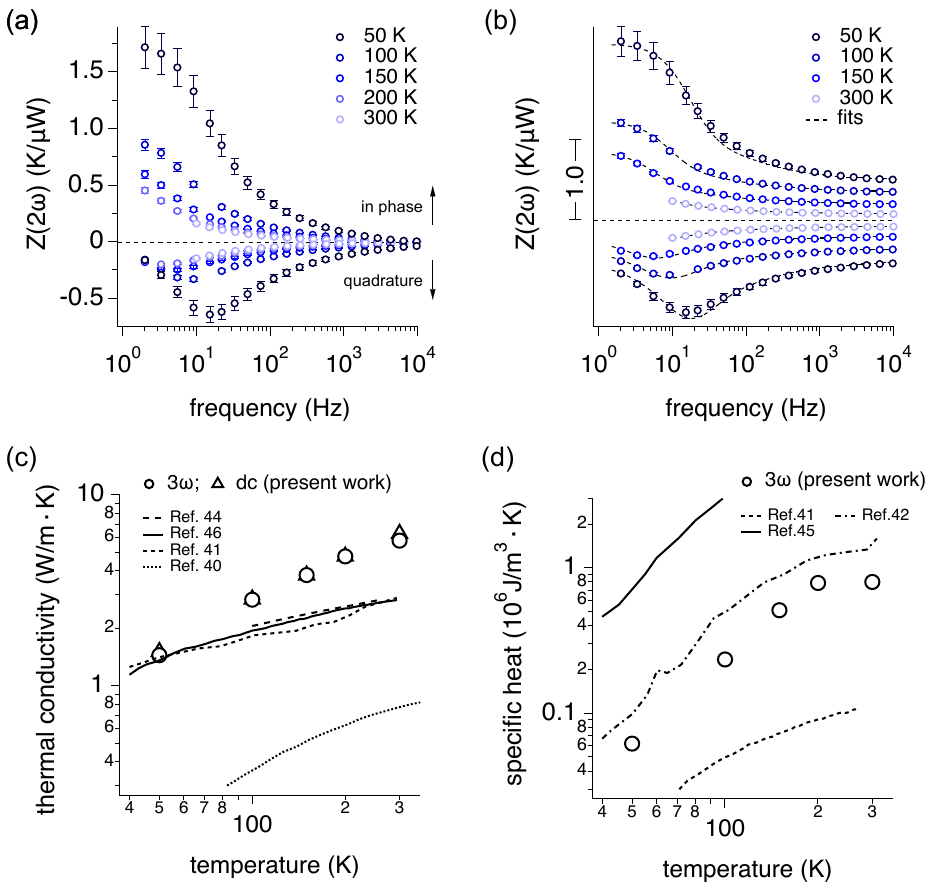}
    \caption{(a) The thermal impedance at the heating frequency, $Z(2\omega)$, vs frequency and temperature. (b) Same as (a), with data fit to Eq.\ \ref{zeq}; 200 K is shown in Fig.\ \ref{3w}. Traces are vertically offset for clarity. (c) $\kappa$ and (d) $c$ for the SiN membrane extracted from the fits to $3\omega$ data (circles) and from separate measurements at dc (triangles; $\kappa$ only.). Error bars are symbol size or smaller. Comparison data are plotted by digitizing prior literature \cite{Lee1997,Zink2004,Queen2009,Sikora2013,Ftouni2013,Ftouni2015}.}\label{z2w}
\end{figure}

In Fig.\ \ref{z2w}(a) we plot the experimentally determined in phase and quadrature values of $Z(2\omega)$, calculated from the $V_{3\omega}$ data of Fig.\ \ref{dev}(b) using Eq.\ \ref{zm}. The error is dominated by uncertainty in $\beta$, which can readily be improved in future experiments. As anticipated, the data all show a qualitative low pass filter behavior, with the ``$RC$'' time decreasing as the temperature is lowered, reflecting a generic decrease in both the thermal conductivity and specific heat. 

In Fig.\ \ref{z2w}(b) we show the results of performing simultaneous fits of the Re and Im parts of Eq.\ \ref{zeq} to the in phase and quadrature data of Fig.\ \ref{z2w}(a) (except for $T=200$ K, shown already in Fig.\ \ref{3w}). With $\kappa$ and $c$ as the only free parameters, we are able to find very good agreement between the expected form of $Z(2\omega)$ and the experimental data, and plot the resulting $\kappa$ and $c$ in Figs.\ \ref{z2w}(c) and (d).

In performing these fits, we explored whether including a parasitic thermal conductance via the Au wire (the $G_\w$ term in Eq.\ \ref{zeq}) or any background gas load ($K_{\perp}$, $c_g$, and $d_g$ in Eqs.\ \ref{keq}-\ref{alphaeq}) made a difference. Using literature values for the thermal conductivity of Au thin films \cite{Mason2020}, we find no difference in either the goodness of fit or the extracted values of the membrane $\kappa$ and $c$ (unsurprisingly, as the heating wire is just one thousandth the width of the silicon nitride membrane). Similarly, any non-zero value for $K_{\perp}$ due to the gas load (the vacuum is $\approx10^{-5}$ Torr in this measurement cryostat) only made the fits worse, and therefore we assume the background gas load contributes negligibly to the thermal transport. Although these effects are not relevant here, we anticipate they may play a role in future work on atomically thin materials, and plan to study thememin in greater detail in future experiments.

In Figs.\ \ref{z2w}(c) and (d) we find our $\kappa$ and $c$ values fall close to---and show a similar temperature dependence as---the results of several prior works \cite{Lee1997,Zink2004,Queen2009,Sikora2012,Sikora2013,Ftouni2013,Ftouni2015}. The comparison data were acquired in silicon nitride grown by a variety of methods which may contribute to the breadth of the results \cite{Zink2004} (for reference, we use SPI part no.\ 4126SN-BA, grown in a low pressure CVD process \cite{Sekimoto1982}), and also use differing experimental approaches. In Fig.\ \ref{z2w}(c), Ref.\ \cite{Lee1997} uses a $3\omega$ measurement tailored for thin films supported on a substrate. The temperature relaxation in a micro- or nano-calorimeter is used in Refs.\ \cite{Zink2004,Queen2009} to access both the thermal conductivity and also the specific heat shown in Fig.\ \ref{z2w}(d). Finally Refs.\ \cite{Sikora2012,Sikora2013,Ftouni2013,Ftouni2015} use a $3\omega$ method similar to our approach, but access the thermal conductivity and specific heat 
via a low-frequency expansion. 

As one means to verify this $3\omega$ approach, we include in Fig.\ \ref{z2w}(c) values of the membrane $\kappa$ measured by applying a dc bias current, using the increase in resistance of the heating wire as a measure of the temperature. Larger currents lead to greater temperatures, from which we extract $\kappa_{dc} = (\Delta P / \Delta T)(l / 2\w d)$, which applies so long as $\Delta T \ll T_0$. These values of $\kappa_{dc}$ closely agree with the $3\omega$ measurements. 

\begin{figure}[t!]
    \includegraphics[width=\columnwidth]{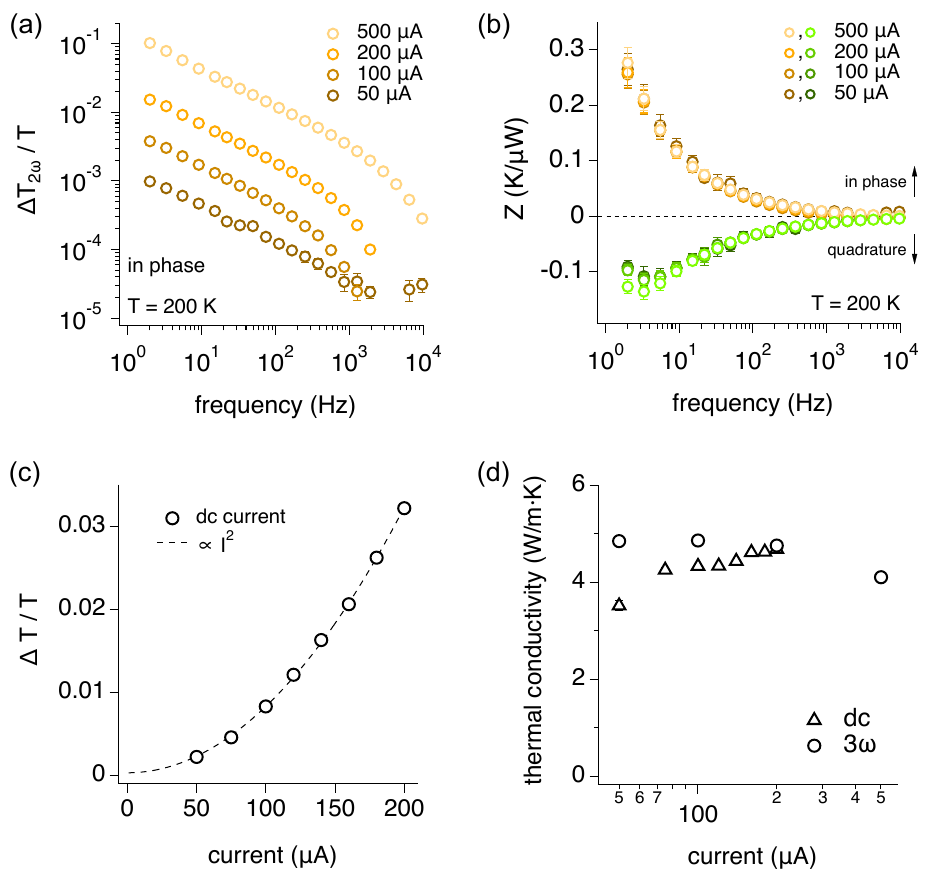}
    \caption{(a) Relative change in temperature, $\Delta T/T$, of the heating wire for various bias currents. (b) in phase and quadrature thermal impedance measured at different temperatures. (c) $\Delta T/T$ induced by a DC heating current. (d) thermal conductivity measured by DC method and AC method.}\label{dT}
\end{figure}

In general it is  desirable to keep the change in temperature small compared to $T_0$. In the $3\omega$ measurements, since $Z$ varies with frequency, the temperature oscillations at $2\omega$ do as well. In Fig.\ \ref{dT}(a) we plot $\Delta T_{2\omega} / T$ for various heating currents vs frequency for the 200 K data. Large temperature swings of up to 10\% at low frequency can be found at the highest current (500 $\mu$A), decreasing to just a part in $10^4$ change for smaller currents at high frequencies. Throughout these variations, Fig.\ \ref{dT}(b) shows the $Z(2\omega)$ values acquired at different bias currents largely agree, with deviations arising only in the peak of the quadrature component at low frequencies for the largest bias. Overall this suggests that keeping $\Delta T/T_0 \lesssim 1\%$ is sufficient.

Figure \ref{dT}(c) shows the change in temperature for the dc bias measurements, calibrated from $R(T)$, has a quadratic increase with current as expected for ohmic heating of the wire. The thermal conductivity shows a weak dependence on bias current for both ac and dc measurements, plotted in Fig.\ \ref{dT}(d). The deviation at low dc bias arises from a small offset voltage in the measurements which is less consequential at higher biases, while the decrease for high ac bias is a consequence of heating beyond the regime of linear response. This results in the dc and ac values agreeing best around $200~\mu$A.

\section{Discussion}

The thermal impedance of a suspended membrane shows a close similarity to a low pass filter. We can define a thermal ``$RC$'' time and a critical frequency $f_c = 1/(2\pi RC)$, which should describe the usual $-3$ dB point of a filter (also the maximum of the quadrature signal). Letting $R = 1/G = l/2 \kappa d \w$ and $C = c d \w l/2$, then $f_c = 2 \kappa / (\pi c l^2)$. For example in Fig.\ \ref{z2w}, the critical values range from $f_c = 4.6$ Hz to 14.6 Hz for the curves from 300 to 50 K, matching the peak location as anticipated. 

This analogy to an electronic low pass filter is intuitively useful but not exact. In particular, the high frequency behavior is $Z_{2\omega}\sim 1/\sqrt{\omega}$ ($-3$ dB/octave), in contrast to $\mathcal{H}\sim 1/\omega$ ($-6$ dB /octave) for the electronic $RC$ filter.

A low frequency or long wavelength limit is often applied in the analysis of $3\omega$ data, such that the thermal wavelength or penetration depth is much larger than the line heater dimensions \cite{cahill_thermal_1990,Sikora2012,Sikora2013}. In this work, the thermal wavelength, $\lambda = 2 \pi / |k| = \sqrt{2 \pi^2 \alpha/\omega}$, is greater than $50~\mu$m for all experimental values of $\alpha \in \{7{\times}10^{-6},2{\times}10^{-5}\}$ m$^2/$s that are found over the frequency range up to $10^4$ Hz, justifying the use of the full form of the impedance in Eq.\ \ref{zeq}.

This method works well so long as parasitic capacitances do not introduce spurious phase shifts. Here, the impedance associated with the distributed cable capacitance (at most 1 nF) remains above 15 k$\Omega$, much greater than the device resistance of 100 $\Omega$. However, we anticipate that if higher frequencies are needed, or if the device resistance is higher, then the $RC$ time $\tau_{RC}$ associated with the distributed capacitance of the electronic cabling and the device resistance will need to be accounted for. There are two clearly relevant cases:\ first, to proceed to lower temperatures $T\lesssim 30$ K in future experiments, materials other than pure metals will be needed in order to have a useful value of $\beta$ (such as AuGe \cite{Scott2022}). These are typically alloys with much larger resistivities, which yield smaller $RC$ times and may impede observation of the peak in the quadrature of the thermal response; in turn, this will limit the accuracy of curve fitting which is most strongly constrained by matching to this peak. Second, in looking ahead to measuring smaller membranes of exfoliated flakes of layered materials (typically 10s of $\mu$m in size), the $1/l^2$ dependence may push the thermal cutoff frequency to approach or exceed the electronic bandwidth, and so require alternative electronic measurement techniques.

Nonetheless, the key advantage of our approach remains:\ that by performing simultaneous fits of the Re and Im parts of Eq.\ \ref{zeq} (or, as needed, the simpler Eq.\ \ref{z0}) to the in phase and quadrature $3\omega$ data, \textit{both} the thermal conductivity and specific heat can be extracted simultaneously for samples in a suspended geometry. Thus the $3\omega$ technique acquires enormous utility for exploring thermal properties in atomically thin materials.

\begin{acknowledgments}
We acknowledge support from and use of the facilities of the Institute of Materials Science and Engineering at Washington University. This work was funded by the Office of the Under Secretary of Defense for Research and Engineering under award number FA9550-22-1-0340, and the Gordon and Betty Moore Foundation under grant DOI 10.37807/gbmf11560.
\end{acknowledgments}

\appendix

\section{Derivation of thermal impedance}

\begin{figure*}[t!]
    \includegraphics[width=\textwidth]{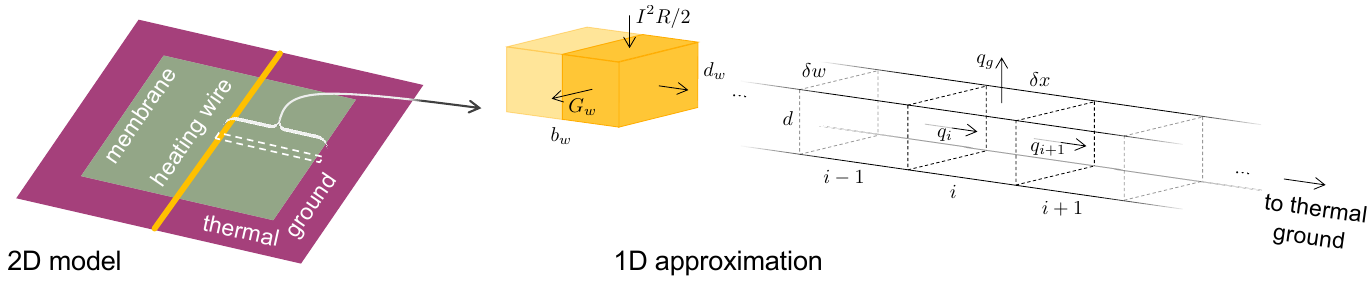}
    \caption{Left:\ 2D model of Fig.\ \ref{3w}, with quasi-1D section highlighted. Right:\ quasi-1D model for calculating heat flow balance in a long uniform rod. Heat is injected by ohmic heating of the metal wire at left (by symmetry only half the membrane and wire is considered), and removed by thermal conductance $G_w$ out the ends of the wire, thermal transport $q_g$ by the presence of any background gas, and by flowing through the rod to thermal ground.}\label{1dmodel}
\end{figure*}

We consider thermal transport in a quasi-1D model of the quasi-2D rectangular membrane, accounting for parasitic heat loss mechanisms including thermal conductance along the heating wire, and heat lost to or gained from the thermal background at temperature $T_b$ (e.g.\ the interior of a cryostat or sample stage) due to the presence of gas in an imperfect vacuum \cite{Nguyen2020}.

If gas is present, we model the membrane as having a thin adsorbed layer with a specific heat $c_g$ and thickness $d_g$; the gas layer is presumed to be in equilibrium with the surroundings such that the adsorption and desorption rates are balanced at a given temperature. Heat may be transported to or from the membrane via a surface thermal conductivity $K_{\perp}$.

In the 1D model, following Fig.\ \ref{3w}, a long rod with thermal conductivity $\kappa$ and specific heat $c$ has heat injected at one end by a thermal mass representing the wire, and the other end is held at the temperature of thermal ground, $T_0$, at the other.

With reference to Fig.\ \ref{1dmodel}, the thermal balance in the $i^{th}$ element of a long rod is given by
\begin{align}
q_i = q_{i+1} + q_g + U_i +U_g
\end{align}
with $q_i = - \kappa A_{\parallel}\delta u / \delta x$ and $A_{\parallel} = d \delta \w$; $q_{g} = 2 K_{\perp}A_{\perp} (u_i-T_0)$ is the gas-mediated surface heat flux from both sides of the membrane, with $A_{\perp} = \delta \w \delta x$ and $u_i$ the temperature of the $i^{th}$ volume element, and for now the background temperature is assumed to be the same as the substrate; and $U_i = c V \delta u_i/\delta t$ and $U_g = c_g V_g \delta u_i / \delta t$ are the internal energies for small volumes $V = d \delta x \delta \w$ of the wire and  $V_g = d_g \delta x \delta \w$ of the adsorbed gas.

Combining terms and taking the continuum limit, the 1D thermal diffusion equation is
\begin{align}
\alpha u_{xx} = u_t + \gamma(u-T_b) \label{diff}
\end{align}
where as above $\alpha = \kappa d / (c d + c_g d_g)$ is the thermal diffusivity,  $\gamma = 2 K_{\perp}/(c d + c_g d_g)$ is a thermal surface conductance, and $u_j$ is the partial derivative with respect to $j$. 

The general solution of Eq.\ \ref{diff} is 
\begin{align}
u(x,t) = 2 A \textrm{sinh}(kx) \textrm{e}^{i2 \omega t} + a \textrm{sinh}(b x) + u_0
\end{align}
with $\omega$ the frequency of the applied heating current. The wavenumber $k=\sqrt{(i 2 \omega + \gamma)/ \alpha}$ and $b=\sqrt{\gamma/\alpha}$\ . The constants $A$, $a$, and $u_0$ are determined by matching to boundary conditions:\ at the cold end of the rod $u(0,t) = T_0$, while at the warm end (at $x=l/2$) the electrical heating power deposited in (half of) the wire per length is balanced against the heat capacity  and the flows of heat into the membrane and out both ends of the wire, expressed as:
\begin{align}
p(t)=c_\w d_\w \frac{b_\w}{2} u_t(l/2,t)\ +\ & \kappa d u_x(l/2,t) \nonumber \\
& + \frac{G_\w}{\w}\left(u(l/2,t)-T_0\right)
\end{align}
Here $p = I^2 \rho / 2 = (I_0^2 \rho/4)(1-\textrm{cos}(2\omega t)) = \textrm{Re}[p_0(1-\textrm{e}^{i 2 \omega t})]$ with $\rho = R/\w$, and other dimensions defined in Fig.\ \ref{3w} and \ref{1dmodel}.

Strictly speaking, in the 1D approximation of Fig.\ \ref{1dmodel}, there is no heat flow to the ends of the heating wire. However the real device is of finite width and the wire is presumed to thermalize to the substrate at $T_0$ at the membrane edges, so the thermal conductance of the wire will remove some heat from the system. To account for this, we derive an effective conductance $G_\w$ for an insulated self-heated wire of finite length $\w$. From the well-known parabolic temperature distribution at zero frequency (which holds even for $\omega \tau > 1$ \cite{dames_1omega_2005}), it is straightforward to find $G_\w = 2 \kappa_\w d_\w b_\w / \w$ for the total effective conductance to both ends of the wire, which ultimately reduces the power available to heat the membrane.

The full solution is
\begin{align}
u(x,t) =&  \frac{-p_0\ \textrm{sinh}(k x)\ e^{i 2 \omega t}}{d k \kappa \textrm{cosh}(k l/2) + (G_\w/\w + i \omega b_\w d_\w c_\w) \textrm{sinh}(k l/2)} \nonumber  \\ 
& + \frac{p_0 \textrm{sinh}(b x)}{d b \kappa \textrm{cosh}(b l/2) + \frac{G_\w}{\w}\textrm{sinh}(b l /2)}+T_0 
\end{align}

We desire to find the impedance at the heating frequency, $Z(2 \omega)$, so that measurements of $V_{3\omega}$ in conjunction with Eq.\ \ref{zm} can be used to obtain $\kappa$ and $c$. Per Eq.\ \ref{conv}, $Z_{2\omega} = \Delta T_{2\omega} / P_{2\omega}$, where the total power deposited in the wire at $2\omega$ is $-p_0\w\textrm{e}^{i 2 \omega t}$. Dividing by this power and simplifying gives the final result shown in Eq.\ \ref{zeq}. A factor of $\nicefrac{1}{2}$ accounts for thermal conduction through two halves of the membrane in parallel.

In cases where the wire negligibly contributes to the thermal conductance and the background gas load can be ignored, these formulae further simplify to
\begin{align}
    Z_0(2\omega) = \frac{1}{2 \w}\left(d k \kappa\ \textrm{coth}(kl/2) + i \omega b_\w c_\w d_\w \right)^{-1} \label{z0}
\end{align} 
with $k=\sqrt{i 2 \omega c/\kappa}$ and $\gamma = 0$.


\bibliographystyle{nsf_erik}

\bibliography{therm.bib}

\end{document}